\documentclass[preprint,12pt]{aastex}
\usepackage{graphics}
\usepackage{amssymb,amsmath}
\usepackage{subfig}

\shorttitle{A Deep {\it Chandra} Observation of 4C~+00.58}
\shortauthors{Hodges-Kluck et al.}

\begin{document}

\slugcomment{Accepted by ApJL}

\title{A Deep {\em Chandra} Observation of the X-shaped Radio
Galaxy 4C~+00.58: A Candidate for Merger-induced Reorientation?}

\author{Edmund~J.~Hodges-Kluck$^{1}$, Christopher~S.~Reynolds$^{1}$, M.
 Coleman Miller$^{1}$ \& Chi.~C.~Cheung$^{2}$} 
\altaffiltext{1}{Department of Astronomy, University of Maryland, College
Park, MD 20742-2421}
\altaffiltext{2}{National Research Council Research Associate, Space
Science Division, Naval Research Laboratory, Washington, DC 20375}
\email{ehodges@astro.umd.edu}

\begin{abstract}
Although rapid reorientation of a black hole spin axis (lasting less than a
few Myr) has been suggested as a mechanism for the formation of wings in 
X-shaped radio galaxies (XRGs), to date no convincing case of reorientation 
has been found in any XRG.  Alternative wing formation
models such as the hydrodynamic backflow models are supported by observed
trends indicating that XRGs form preferentially with jets aligned along the
major axis of the surrounding medium and wings along the minor axis.  
In this Letter, we present a deep {\em Chandra} observation of 4C~+00.58, 
an oddball XRG with its jet oriented along the {\em minor} axis.  By using
the X-ray data in tandem with available radio and optical data, we
estimate relevant timescales with which to evaluate wing formation models.
The hydrodynamic models have difficulty explaining the long wings, whereas
the presence of X-ray cavities (suggesting jet activity along a prior axis)
and a potential stellar shell (indicating a recent merger) favor a
merger-induced reorientation model.  
\end{abstract}


\section{Introduction}
X-shaped radio galaxies (XRGs) are double-lobed radio galaxies \citep{leahy84} 
which also possess a pair of long, faint, centro-symmetric ``wings''.  
They have gained notoriety as a possible signature of a rapid
(within a few Myr) reorientation of the supermassive black hole (SMBH) spin axis,
conceivably induced by galaxy mergers in which either accretion torque or
a SMBH merger causes a spin-flip \citep{rottmann01,merritt02}.  In this 
scenario, wings are ``fossils'' tracing the prior jet axis which radiatively
decay.

However, no convincing case for reorientation has been made in any individual
XRG, whereas several lines of evidence support a hydrodynamic origin for the
wings.  For instance, in most XRGs, the wings are coaligned with the minor axis of the
host galaxy and the jet with the major axis \citep{capetti02,saripalli09}.
A similar major-axis---radio alignment trend exists in the X-rays
\citep{hodges10}.  These results have been interpreted to favor models in
which the wings are produced by radio-lobe---gas interaction
\citep{leahy84,worrall95,capetti02}.  Additionally, no clear signs of mergers have been
found in a spectroscopic study of XRG hosts, whereas they may be
overpressured \citep{landt10}.  

In this Letter, we identify 4C~+00.58 \citep[Fig.~1, classified as a candidate
XRG by][]{cheung07} as one of the best candidates for a merger-induced
reorientation based on quantities derived from a deep {\it Chandra X-ray Observatory} observation and 
publicly available data.  Unlike other XRGs, the 4C~+00.58 jet is coaligned 
with the {\it minor} axis of its host. 
Even if only a fraction of XRGs are produced by reorientation, their
frequency may be important for estimating {\it Laser Interferometer Space
Antenna} detection rates. 

We use a Galactic absorption of 
$N_H = 7.14\times10^{20}$~cm$^{-2}$ \citep{kalberla05}, 
as well as the {\em Wilkinson Microwave Anisotropy Probe} cosmology 
\citep[$H_0 = 71$~km~s$^{-1}$~Mpc$^{-1}$, $\Omega_{\Lambda} = 0.73$, and 
$\Omega_m = 0.27$;][]{spergel07}.  At a redshift of $z = 0.059$, 
$1^{\prime\prime} = 1.13$~kpc. 

\section{Observations}

We obtained a 93~ks {\em Chandra} exposure toward 4C~+00.58 using the Advanced CCD Imaging 
Spectrometer\footnote{See http://cxc.harvard.edu/proposer/POG/pdf/ACIS.pdf} 
(ACIS) and combined it with a 10~ks archival observation 
\citep[obs. IDs 10304 and 9274; latter published in][]{hodges10}.  
The source is centered at the nominal aim point on the ACIS-S3 chip.  
The data were reduced with
the {\em Chandra} Interactive Analysis of Observations (CIAO v4.0) software,
and spectral fitting was performed with XSPEC \citep{arnaud96}.
We extracted a $0.3-10$~keV bandpass lightcurve (binned to 600~s) from empty
regions to check for background flares, but found no 3$\sigma$ deviations.  
The extended emission around 4C~+00.58 is less than 45$^{\prime\prime}$ in 
radius, so we use local background for spectral extraction. 

We also use a NRAO\footnote{The National Radio Astronomy Observatory 
is a facility of the National Science Foundation operated under cooperative
agreement by Associated Universities, Inc.} {\em Very Large Array}
\citep[VLA;][]{thompson80} 1.4 GHz map \citep{hodges10} and a 4.9~GHz map
produced by combining archival snapshot A-array data from \citet{best99} and
C-array data from program AC406.  We use Sloan Digital Sky Survey \citep[SDSS;][]{adelman08} red (623.1~nm) and green
(477.0~nm) images from two 54~s exposures of the host 
(SDSS J160612.68+000027.1) to measure color and magnitude,
correcting for the smaller point spread function of the green images as well
as sky background, the 1000-count software bias, and Galactic extinction.

The 1.4~GHz map is shown in the left panel of Figure~1.  The
primary lobes of the radio galaxy lie nearly on an east--west axis and
have a well defined boundary, whereas the faint wings are
oriented in a north--south direction.  The jet experiences a dramatic bend
(by 60$^{\circ}$) just before terminating, and the cocoon in the 1.4~GHz
map is notable for a well defined edge with a surface brightness about
five times that of the wings.  The 5~GHz map resolves the jet into a
string of knots (\S3.3).  We detect no counterjet.

The X-ray emission (Fig.~1) is made up of two components: bright
emission spatially associated with the jet/AGN and a compact diffuse
atmosphere.  To isolate the atmosphere, we mask point-like sources and
restrict the energy bandpass to $0.3-3$~keV (Fig.~2); this energy range
contains 80\% of the photons within 45$^{\prime\prime}$.  On the basis of
an unsharp mask image and a weighted Voronoi tessellation
adapatively binned image \citep{diehl06,cappellari03}, we have identified
several X-ray cavities (Fig.~2).  These cavities, labeled C$_{\text{NE}}$, 
C$_{\text{NW}}$, C$_{\text{SE}}$, and C$_{\text{SW}}$, are low surface
brightness regions in the hot atmosphere bounded by ``spurs'' of greater
surface brightness.  The cavities are deep negatives in an
unsharp mask image in which a heavily smoothed map (40~px) is subtracted
from a lightly smoothed (3~px) map.  The residuals are shown in Fig.~2.

The SDSS image (Fig.~3) reveals a dim, resolved extension to the 
southwest of the host galaxy.  The elliptical host shows
no obvious internal structure, but the extension may be a stellar
shell from a prior minor merger \citep[for a discussion of shells see][]{quinn84}.
The extension is red $(g-r \sim 0.7)$, but bluer than the host $(g-r \sim 1.0)$.
The apparent magnitude of the entire extension is 
$m_r \sim 20$.

\section{Results}

\subsection{Hot Atmosphere}

The bright X-ray emission immediately near the jet is nonthermal, otherwise the
extended emission is consistent with isothermal plasma out to 50~kpc from
the AGN, with $kT = 1.1\pm 0.2$~keV within 25~kpc and $kT = 0.9\pm0.1$~keV
outside.  The photon statistics preclude a deprojection analysis.
On the basis of the resulting emission measures, densities within 25~kpc are a 
few$\times 10^{-3}$~cm$^{-3}$, implying pressures $P = 10^{-12}$ to 
$10^{-11}$~dyne~cm$^{-2}$.  Thus, we adopt a sound speed 
$c_s \approx 400$~km~s$^{-1}$ throughout the region.  
The total luminosity of the hot atmosphere is 
$L_X \sim 3 \times 10^{41}$~erg~s$^{-1}$.

The gross morphology of the X-ray emission does not coincide with the host
galaxy.  Using only the 10~ks exposure, we argued \citep{hodges10} that the
orientation agreed with the host although the ellipticity did not.  With
deeper data it is evident that while we correctly excluded nonthermal
emission to the southwest, we could not exclude nonthermal emission to
the northeast. 

The most notable feature of the diffuse X-ray map is C$_{\text{NW}}$ (Fig.~2),
which is collinear with C$_{\text{SE}}$ and the AGN.  C$_{\text{NW}}$ is
enclosed by emission and cospatial with a spur in the radio cocoon, suggesting
that C$_{\text{NW}}$ and C$_{\text{SE}}$ are jet-blown cavities.  
C$_{\text{NE}}$ and C$_{\text{SW}}$ are also collinear with the AGN and are
associated with the bases of the wings, with walls extending into the 
surrounding medium. 

The free-free cooling time of the C$_{\text{NW}}$ and C$_{\text{SE}}$ bounding 
material places an upper limit on their ages: 
\begin{equation}
t_{\text{ff}} \sim \frac{5}{2}\frac{nkT}{\Lambda_{\text{ff}}(T)} \approx 1.6 \times 10^{9} 
T_7^{1/2} n_{-3}^{-1} \text{\hspace{0.2cm}yr} \sim 500 \text{\hspace{0.2cm}Myr}
\end{equation}
where $T_7 \sim 1.4$ is the temperature in units of $10^{7}$~K and $n_{-3} \sim 4$ is the 
density in units of $10^{-3}$~cm$^{-3}$.  If C$_{\text{NW}}$ is a
spherical bubble (of radius 5~kpc) inflated at a locally estimated
pressure
$P = 5\times 10^{-12}$~dyne~cm$^{-2}$, the work done to inflate it is
$W \sim 7\times 10^{55}$~erg.  Since the transonic expansion time is 10~Myr, 
the minimum average kinetic luminosity of the jet during that period is 
$2\times 10^{41}$~erg~s$^{-1}$ if C$_{\text{NW}}$ is jet-blown.

\subsection{Wings}

The long wings are associated with low-signal X-ray structure (Fig.~2, bottom
right).  The approximate wing symmetry implies coherent formation, but the 
southern wing is slightly longer.

If the wings expanded transonically, their length ($\sim 36$~kpc) implies an
age of 90~Myr, although this is a minimum age since the wings are seen in
projection and may expand subsonically.  Conversely, the synchrotron
decay time $t_{\text{sync}}$ provides a maximum wing lifetime 
assuming the radio emission traces the entire wing volume and
the wings were inflated by the radiating plasma.  To estimate
$t_{\text{sync}}$, we follow \citet{tavecchio06} and use the 1.4~GHz map to
estimate the equipartition field $B_{\text{eq}}$.  We take the spectral index,
$\alpha = 0.7$ ($\S_{\nu} \propto \nu^{-\alpha}$), 
from low-resolution radio photometry and assume
$\gamma_{\text{min}} \sim 10$.  For cylindrical wings of $r = 6$~kpc and
$h = 36$~kpc, we obtain $B_{\text{eq}} \sim 10$~$\mu$G.  We then find the
electron Lorentz factor $\gamma \sim 5600$ from  
$\nu_{s\text{[1.4 GHz]}} = 4\times 10^{-3} B \gamma^2 = 1.4$~GHz, and thus find
\begin{equation}
t_{\text{sync}} \approx 2.4\times 10^{9} \gamma_4^{-1} B_{\mu{\text G}}^{-2} \text{ yr} \sim 40 \text{\hspace{0.2cm}Myr}
\end{equation}
where $\gamma_4$ is in units of $10^4$ and $B_{\mu\text{G}}$
is in $\mu$G.  This value represents the cooling time of the 
1.4~GHz electrons; we emphasize our assumption that this is ``first-generation''
plasma occupying the wings.  $t_{\text{sync}}$ is insensitive to projection
effects relative to the transonic expansion time: if the wings are longer
by a factor of 2, $B_{\text{eq}}$ decreases by a factor 
$2^{1/(3+\alpha)} \sim 1.2$ and $t_{\text{sync}}$ increases by a factor
$\sim 1.4$.  The disagreement between $t_{\text{sync}}$ and the expansion time 
suggests either supersonic expansion (i.e. like a jet-blown cocoon) or 
wing replenishment by supersonic inflowing lobe plasma. 

The work required to inflate the wings, assuming the cylinders above, 
is $PdV \sim 10^{57}$~erg.  An age of 40~Myr implies an average kinetic 
luminosity of $L_{\text{kin}} \sim 8\times 10^{41}$~erg~s$^{-1}$ applied to the 
wings alone.  

\subsection{Nucleus and Jet}

The strong central X-ray point source corresponds to the AGN and contains 1100
counts.  The spectrum is fit well by a model
consisting of an unabsorbed power law with spectral index $\alpha = 0.7\pm0.1$ and
a weak thermal model frozen at $kT = 1.0$~keV.
The nonthermal luminosity is $L_X \sim 7\times10^{41}$~erg~s$^{-1}$ between 
$0.3-10$~keV.

The X-ray jet is the next brightest feature and traces the radio jet well,
including several X-ray knots (Fig.~4).  A super-sampled X-ray image
reveals that two of these knots coincide with more compact knots visible in
the 5~GHz image and a bright region in the 1.4~GHz jet.  There is a modest
radio ($1.4-5$~GHz) spectral gradient between the inner and outer jet (Table~1), and
the outer jet does not appear in the X-rays.  Along the inner jet, we measure
a broad-band radio to X-ray spectral index, $\alpha_{rx} \sim 1.0$, which
is consistent with the X-ray spectrum
($\alpha_x \sim 1.1\pm0.4$).  Thus, the X-ray emission is consistent with
a synchrotron origin (requiring a concave-down spectral energy distribution).  
Alternatively, assuming an
inverse Compton origin for the X-rays, we follow \citet{tavecchio06} to
estimate the jet Doppler factor $\delta = [\gamma(1-\beta \cos\theta)]^{-1}$
in the knots by finding $B$ such that 
$B_{\text{eq}} \delta= B_{\text{IC}}/\delta$.  This
results in $\delta \sim 10$ (Table~1) and $B_{\text{knot}} \sim 10$~$\mu$G.
As the jet X-ray emission is unlikely dominated by inverse-Compton emission,
$\delta$ is formally an upper limit.  
$B_{\text{eq}}$ is derived with no knowledge of the X-ray emission 
and predicts $P \sim 3\times 10^{-12}$~dyne~cm$^{-2}$, in agreement with 
spectral fitting estimate.  

Although no counterjet is visible, the 1.4~GHz emission 
suggests the jet has point symmetry.  If the jet tail is dragged, the tail 
will cool radiatively with its length set by the cooling time (i.e. no
additional plasma influx).  
We derive an upper $B_{\text{eq}}\delta \sim 120$~$\mu$G at $\nu_s = 5$~GHz
and thus find $t_{\text{sync}} \sim 0.5 - 15$~Myr for $\delta = 1 - 10$.  
We assume $\delta$ declines along the jet, so for $\delta \sim 5$, 
$t_{\text{sync}} \sim 6$~Myr. 

The projected length of the radio cocoon gives a transonic expansion time 
of 35~Myr.  The work done to excavate a cocoon with semimajor 
axis $a = 17$~kpc and semiminor axes $b=c=8$~kpc is 
$PdV \sim 7\times 10^{56}$~erg, so the minimum average
kinetic luminosity while inflating the cocoon is 
$L_{\text{kin}} \sim 7 \times 10^{40}$~erg~s$^{-1}$.

\section{Wing Formation Models}

We consider three wing formation scenarios: an overpressured outburst, 
conical precession of the jet axis, and merger-induced reorientation of the jet 
axis.  

In the backflow models, wings are produced by pressure- or buoyancy-driven 
back-flowing plasma from the terminal shocks evolving in the hot medium.  Since we see
no obvious evidence for a plumed jet {\it directly} feeding 
the wings, the most plausible of these models is a ``blow-out'' from an 
overpressured cocoon early in the source's life \citep{capetti02}.  In this 
model, the native atmosphere confining the young jets is aspherical
with a preferential direction along which the cocoon ruptures. 

The backflow model suffers from several difficulties.  First, the
wings must expand at most transonically.  The long projected length of the wings in 
4C~+00.58 (requiring an AGN lifetime of at least 90~Myr) is difficult to 
reconcile with the cocoon length unless the cocoon
inclination angle from the line of sight $\theta_{\text{LOS}}$ is less than
$30^{\circ}$.  The equivalent widths of \ion{O}{2} (3727\AA)
and \ion{O}{3} (5007\AA) measured by \citet{landt10} argue against a steep
inclination \citep{landt04}. 
Furthermore, the maximum wing lifetime $t_{\text{sync}}$ 
disagrees with the transonic expansion time, although the wing plasma may
be continuously replenished.  Second, the model relies on strong backflows
typically associated with \citet[][FR]{fan74} type~II sources. 
4C~+00.58 is not easily classified as FR~I or II, but at $M_R \sim -22.7$ and
$\log{L_r(\text{1.4 GHz})} \sim 25.3$~W~Hz$^{-1}$ falls very close to the
\citet{ledlow96} boundary ($L_{\text{radio}} \propto L_{\text{opt}}^{1.8}$).
Since XRGs generally lie near the \citet{ledlow96} FR~I/II boundary, they
may be a transition population \citep{cheung09,landt10}.  
Finally, the radio galaxy is misaligned with its host,
so the \citet{capetti02} model cannot produce wings until the jet escapes the
interstellar medium.  

Conical precession is a simple model in which the jet axis swings around, so
the wing extensions and post-bend jet are equidistant from the AGN.  
This requires a steep inclination angle
of $\theta_{\text{LOS}} < 30^{\circ}$.  In this model, the wings trace the jet
history, and the oldest plasma can be no older than the synchrotron
cooling time.  To obtain the precession rate $\dot{\phi}$, we compare the
projected length of the post-bend jet to its cooling time (about 6 Myr) and
obtain $\dot{\phi} \sim 4^{\circ}$~Myr$^{-1}$ (corresponding to 
a mildly supersonic $v \sim 460$~km~s$^{-1}$).  A 180$^{\circ}$ 
rotation takes $\sim 45$~Myr and implies supersonic expansion for the 
cocoon.  The timescale is consistent with the wing $t_{\text{sync}}$, but
the model must also explain the larger far-side cone, as both the far-side
cocoon and wing have longer projected lengths.  Notably, precession does not
explain the presence of C$_{\text{NW}}$ and C$_{\text{SE}}$, and 
numerical simulations \citep[e.g.][]{falc10} indicate that it
would not preserve obvious cocoon structure. 

In the reorientation (spin-flip) scenario, the wings are fossil lobes of a
jet whose direction rapidly changed, either due to accretion torque or
coalescence of a SMBH binary.  There is circumstantial evidence for such a 
spin-flip: a possible stellar shell indicating a minor merger, and the
cavities C$_{\text{NW}}$ and C$_{\text{SE}}$ with overlapping cocoon
extensions implying somewhat recent jet--gas interaction along an 
old axis.  We describe a reorientation scenario for 4C~+00.58 presently.

Given the small size of C$_{\text{NW}}$ and C$_{\text{SE}}$, the jet was in 
a weak or ``off'' state prior to the minor merger, but the SMBH spin axis was 
aligned with the major
axis of the host.  Upon ignition, the jet quickly formed C$_{\text{NW}}$ 
and C$_{\text{SE}}$.  However, since the angular momentum axis of the accreting
gas is generally misaligned with that of the SMBH, accretion torque will 
reorient the black hole's spin within a few Myr \citep{dotti09}.  Hence,
C$_{\text{NW}}$ must have expanded at $v_{\text{exp}} > 2c_s$ if no prior
cavity existed.

Once accretion torque moved the jet to the wing axis, it inflated the wings as
active lobes.  The transonic lateral expansion time of the wings 
is 10~Myr, implying $v_{\text{exp}} < 8c_s$ during wing inflation.  
The ``Z-shaped'' wing extensions (covering 30$^{\circ}$) could be 
explained either as post-reorientation ``wiggles'' from a hot disk 
\citep[][show that a hot disk is required for wiggles of this magnitude]{dotti09}
or interaction between the lobes and merging ISM swirling into the host
\citep{gopal03,zier05}.  These Z-shaped extensions
then evolve buoyantly, and may be replenished by the primary lobes once the jet axis
has moved \citep{gopal03}.  We suppose the jet moved to its current position
due to coalescence of the SMBH binary or ongoing accretion torque, then formed
the present cocoon.  Since the jet may have experienced small 
realignments during the wing inflation phase, we infer a reorientation 
timescale of fewer than 50~Myr since jet ignition, well within the 
free-free cooling time of the
C$_{\text{NW}}$ walls and constrained by the wing $t_{\text{sync}}$.  
Although it is possible that the system represents only a single spin-flip
from the wings to the present location, this hypothesis does not not explain
C$_{\text{NW}}$ or C$_{\text{SE}}$. 

The presence of a stellar shell must be confirmed, and it is possible that
C$_{\text{NW}}$ and C$_{\text{SE}}$ are not jet-blown cavities 
but rather part of a cocoon-evacuated shell (with bounding material describing
a ring perpendicular to the jet) produced by a 
jet-ignition shockwave.  The overlap of C$_{\text{NW}}$ by
the radio cocoon is then due to backflow filling the cavity.  
Assuming a circular ring, the eccentricity of the ring implies 
$\theta_{\text{LOS}} \sim 60^{\circ}$, far above the 30$^{\circ}$ required
to reconcile the cocoon and wing lengths. 

\section{Context and Summary}

There are few deep X-ray observations of XRGs.  Apart from 4C~+00.58, there is
a $\sim 100$~ks {\em Chandra} observation of NGC~326 and a 50~ks exposure of 
3C~403 \citep{kraft05}.  These data show X-ray emission on
different scales and of differing surface brightness, forcing the hydrodynamic
hypothesis to contend with a variety of environments.  The cavities in
4C~+00.58 also demonstrate that X-ray observations are not only useful for 
studying backflow models.  Our prior survey \citep{hodges10} and this study
suggest exposure times of at least 100~ks are required to examine
detailed structure. 

We know of no clean evidence for merger-induced reorientation.  Even
in our toy model, the wings are produced by merger-induced accretion rather
than an instantaneous spin-flip, so the black hole merger itself would 
involve mostly-aligned spins \citep{bogdanovic07}.  Nonetheless, the 
presence of an apparent stellar shell suggests that searching for structure
in the hosts of XRGs may provide strong indirect evidence for such mergers
in a subclass of these objects.

We have presented a deep {\em Chandra} observation of the XRG 4C~+00.58.  
The hot atmosphere is roughly co-spatial with the radio galaxy and has a 
temperature of $kT \approx 1.0\pm0.2$~keV.  An X-ray jet of about 5~kpc is
detected, overlapping well with the 5~GHz knots.  We synthesize information 
from the radio and X-ray maps to assess 
three wing-formation models based on approximate limiting timescales and argue that
the hydrodynamic scenario faces several difficulties whereas circumstantial
evidence favors the reorientation model.  Although 4C~+00.58 does not obey
the optical--radio correlation of XRGs \citep{capetti02,saripalli09}, 
confirmation of any of the wing formation scenarios would bear on XRG 
formation generally.

\acknowledgments

We thank the referee for helpful suggestions and clarification.  
E.~H.-K. and C.~S.~R. thank the support of the {\em Chandra} Guest Observer 
program grants GO90111X and GO89109X.  

{\it Facilities:} \facility{CXO}, \facility{VLA}

%% Figures and Tables

\clearpage
%% Table 1: X-ray knots 

\begin{deluxetable}{lccccccccc}
\tablenum{1} 
\tabletypesize{\scriptsize} 
\tablecaption{Core and Jet Parameters} 
\tablewidth{0pt} 
\tablehead{ 
\colhead{Designation} & \colhead{Distance} & \colhead{$F_{\text{1.4 GHz}}$} &
\colhead{$F_{\text{4.9 GHz}}$} & \colhead{$F_{\text{X-ray}}$} & 
\colhead{$\alpha_r$} & \colhead{$\alpha_{rx}$} & \colhead{$\alpha_x$} & 
\colhead{$\delta$} & \colhead{$B_{eq}$} \\
& ($^{\prime\prime}$) & (mJy) & (mJy) & (nJy) &  &  &  &  & ($\mu$G)}
\startdata
Core          & 0.0       & 41$\pm$4   & 30$\pm$3  & $8.3^{+0.7}_{-0.1}$  & 0.25                 & 0.8 & 0.7$\pm$0.1 & -  & - \\
Inner jet     & 0.6$-$6.0 & 180$\pm$20 & 88$\pm$9  & $1.0\pm0.3$  & 0.6                  & 1.0 & 1.1$\pm$0.4 & $<7$  & 7  \\
Inner Knot 1  & 2.5       & -          & 14$\pm$1  & $0.3\pm0.2$  & 0.6\tablenotemark{a} & 1.0 & 1.0$\pm$0.7 & $<8$  & 9  \\
Inner Knot 2  & 3.3       & -          & 15$\pm$1  & $0.9\pm0.3$  & 0.6\tablenotemark{a} & 0.9 & 1.5$\pm$0.6 & $<10$ & 7  \\
Outer jet     & 6.0$-$10. & 220$\pm$20 & 70$\pm$7  & $<0.1$       & 0.9                  & $>$1.1   & -           & -  & 120/$\delta$ \\
\enddata
\tablenotetext{a}{It is not possible to separate the knots at 1.4~GHz, so we use the average value instead of
measuring a flux.}
\tablecomments{\label{jet_table}Most of the jet
X-ray emission comes from knot 2 and the outer jet has no X-ray emission.  The
distance is measured radially from the core in arcsec.  We report a model 
X-ray flux from the best-fit power-law model with errors
reported at 90\% confidence.}
\end{deluxetable}

%\clearpage
%% Figure 1: Raw Images
\begin{figure}
\begin{center}
\includegraphics[scale=0.8]{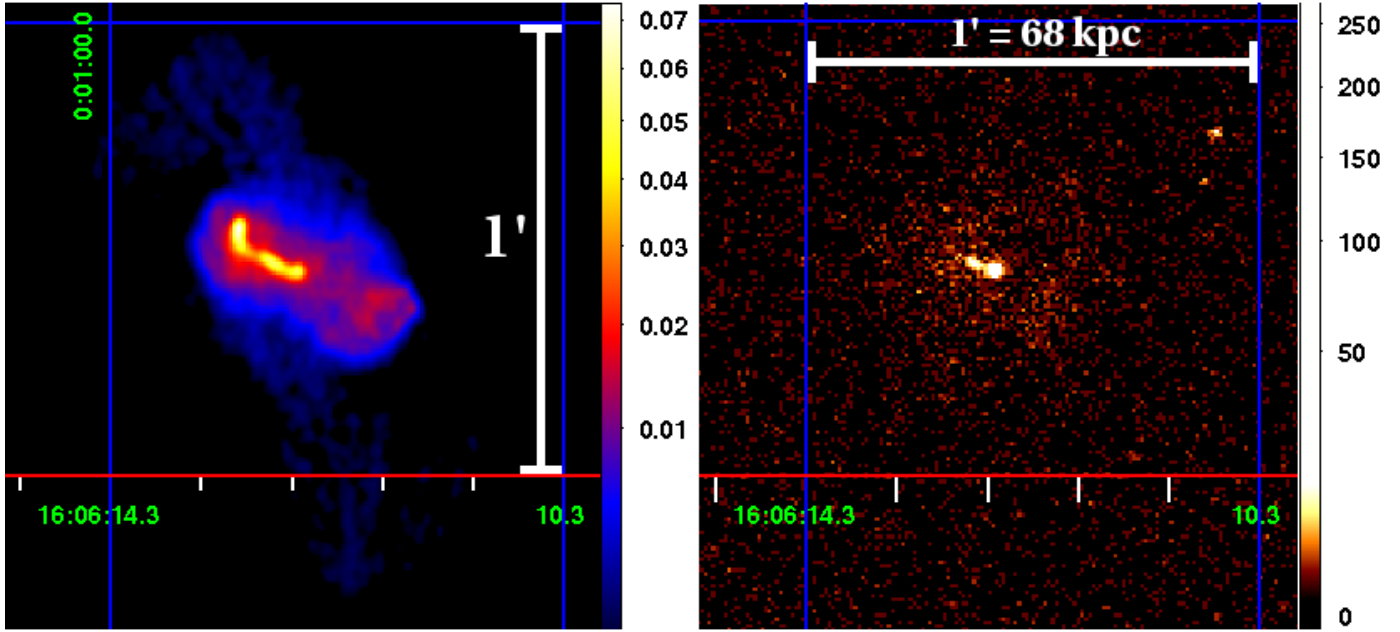}
\caption{{\sc Left:} 1.4~GHz VLA A-array map 
($1.^{\prime\prime}6 \times 1.^{\prime\prime}3$ beam, units in Jy/beam).  
  {\sc Right:} Raw {\em Chandra} image
  ($0.3-10$~keV).  The brightest pixel in the AGN has 282 counts. 
}
\label{images}
\end{center}
\end{figure}

%\clearpage
%% Figure 2: X-ray cavities
\begin{figure}
\begin{center}
\includegraphics[scale=0.65]{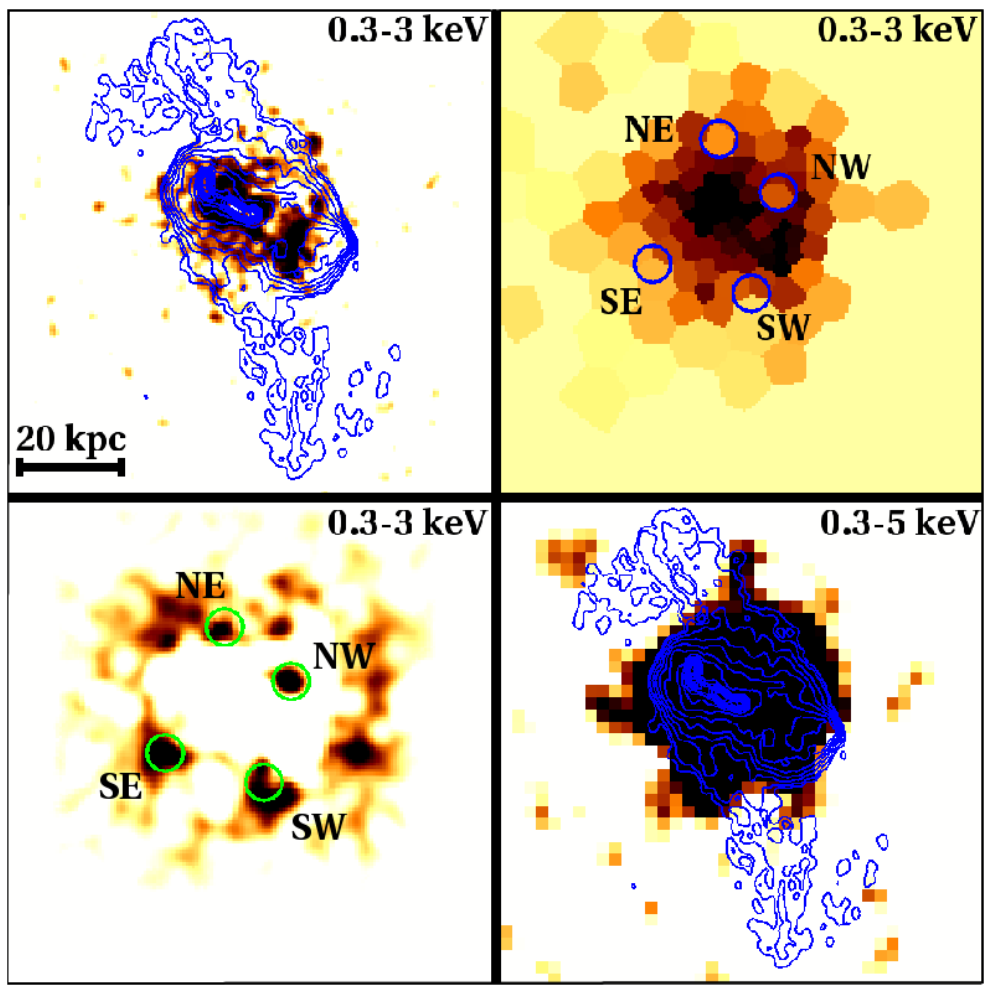}
\caption{{\sc Top Left:} 1.4~GHz contours overlaid on smoothed ($\sigma = 3$~pixels)
X-ray image with point-like sources excised, clipped at twice the mean 
background.  {\sc Top Right:} A weighted Voronoi tessellation image 
(S/N = 5.0 in each tile) of the $0.3-3$~keV events with cavities identified.
{\sc Bottom Left:} Negative of an unsharp mask image of top left (using a 
smoothing length of 40~px for subtraction) showing cavities.  
{\sc Bottom Right:} Coarsely-binned (4$\times$ native pixels) image from $0.3-5$~keV
showing extended structures beyond 20 kpc from the AGN.}
\label{xray_smoothed}
\end{center}
\end{figure}

\clearpage

%% Figure 3: SDSS
\begin{figure}
\begin{center}
\includegraphics[scale=0.85]{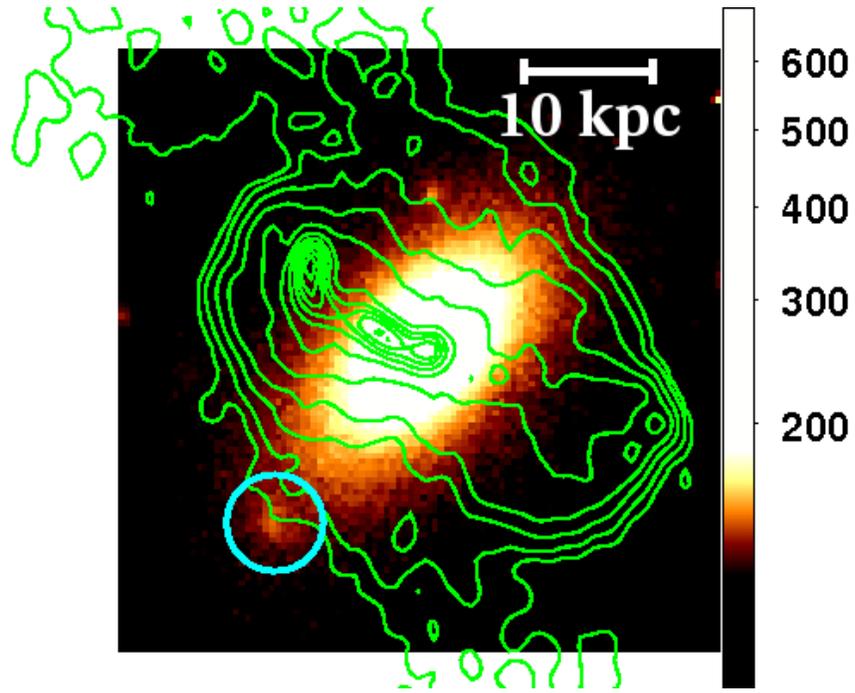}
\caption{1.46~GHz contours overlaid on a combined red$+$green SDSS image
of the host galaxy of 4C~+00.58.  The cyan circle encloses a 20th magnitude
resolved ``extension'' to the elliptical galaxy which is slightly bluer.  
}
\label{sdss}
\end{center}
\end{figure}

\clearpage

%% Figure 4: Jet knots
\begin{figure}
\begin{center}
\includegraphics[scale=0.85]{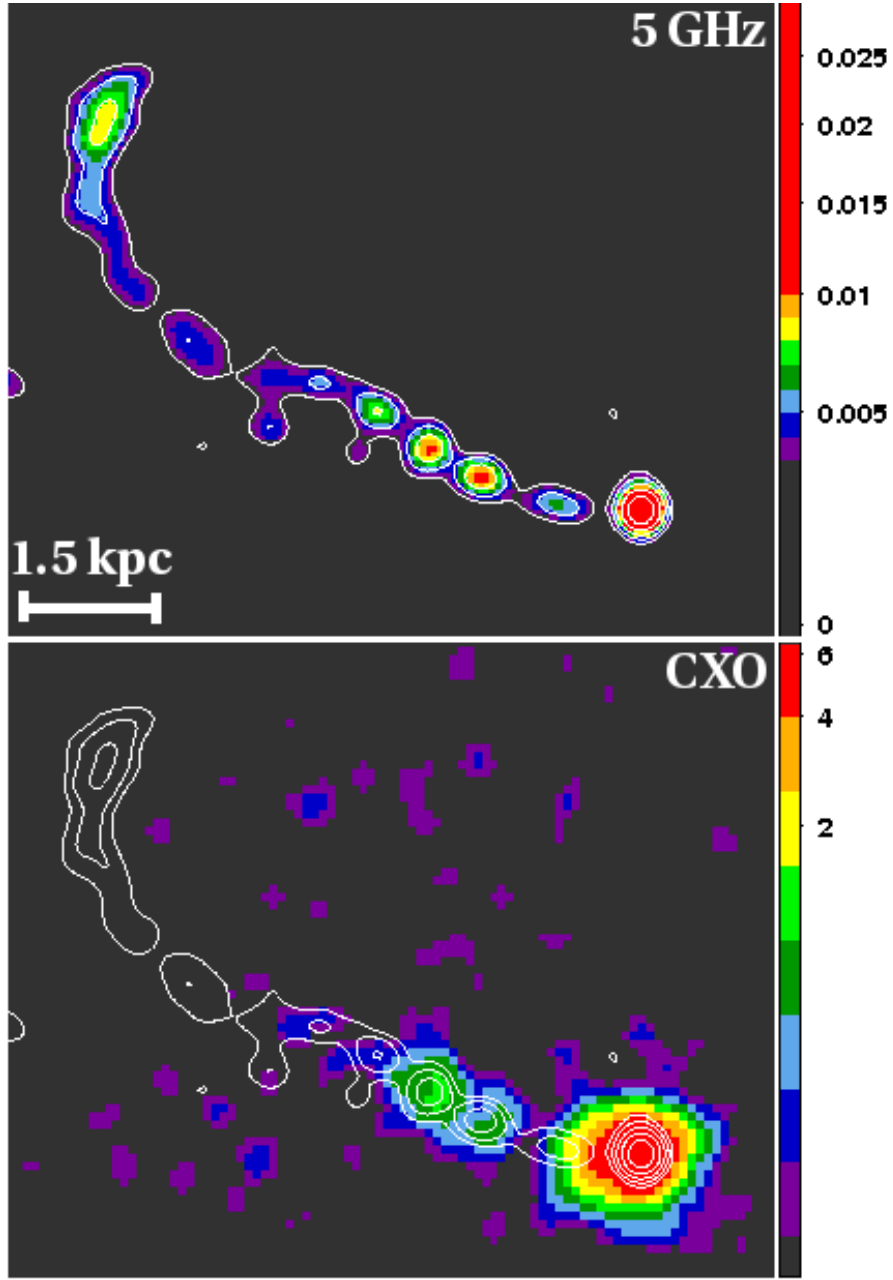}
\caption{{\sc Top:} 5~GHz VLA image of the jet with contours (beamsize
$0.5\times0.5^{\prime\prime}$, units in Jy/beam).  {\sc Bottom:}
Smoothed X-ray image with pixel randomization turned off (superbinned to 
1/4-original pixel size) with 5~GHz contours overlaid.
}
\label{jet_knots}
\end{center}
\end{figure}

\end{document}